\documentclass[qsecnum,amsmath,preprintnumbers,superscriptaddress,nofootinbib,aps,prd,10pt,a4paper,twocolumn]{revtex4-1}
\pdfoutput=1
\usepackage[utf8]{inputenc}
\usepackage[english]{babel}
\usepackage{amsmath}
\usepackage{amsfonts}
\usepackage{amssymb}
\usepackage{amsthm}
\usepackage{hyperref}
\usepackage{color}
\usepackage{tikz}
\usepackage{feynmp-auto}
\usepackage{soul}
\usepackage{bm}
\usepackage{graphicx}
\usetikzlibrary{matrix}

\usepackage{caption}
\usepackage{subcaption}
\usepackage{placeins}

 \def\be{\begin{equation}}
\def\ee{\end{equation}}
 \def\ba{\begin{align}}
\def\ea{\end{align}}
\def\bea{\begin{eqnarray}}
\def\eea{\end{eqnarray}}
\def\d{\partial}

\def\g{\gamma}
\def\m{\mu}

\def\l{\lambda}

\begin{document}
\preprint{INR-TH-2021-004}
\title{{\bf Black Holes in Ultraviolet-Complete Ho\v rava Gravity}}

\author{Guillermo Lara}
\email[]{jlaradel@sissa.it}
\address{SISSA, Via Bonomea 265, 34136 Trieste, Italy and INFN Sezione di Trieste}
\address{IFPU - Institute for Fundamental Physics of the Universe \\Via Beirut 2, 34014 Trieste, Italy}

\author{Mario Herrero-Valea}
\email[]{mherrero@sissa.it}
\address{SISSA, Via Bonomea 265, 34136 Trieste, Italy and INFN Sezione di Trieste}
\address{IFPU - Institute for Fundamental Physics of the Universe \\Via Beirut 2, 34014 Trieste, Italy}

\author{Enrico Barausse}
\email[]{barausse@sissa.it}
\address{SISSA, Via Bonomea 265, 34136 Trieste, Italy and INFN Sezione di Trieste}
\address{IFPU - Institute for Fundamental Physics of the Universe \\Via Beirut 2, 34014 Trieste, Italy}

\author{Sergey M. Sibiryakov}
\email[]{ssibiryakov@perimeterinstitute.ca}
\address{Department of Physics \& Astronomy, McMaster University,\\ Hamilton, Ontario, L8S 4M1, Canada}
\address{Perimeter Institute for Theoretical Physics, Waterloo,
  Ontario, N2L 2Y5, Canada}
\address{Institute for Nuclear Research of the Russian Academy of Sciences
60th October Anniversary Prospect, 7a, 117312 Moscow, Russia}

\begin{abstract}
 Ho\v rava gravity is a proposal for completing general relativity in the
ultraviolet by interactions that violate Lorentz invariance at very
high energies. We focus on (2+1)-dimensional projectable Ho\v rava
gravity, a theory which is renormalizable and perturbatively 
ultraviolet-complete, enjoying an asymptotically free ultraviolet fixed point.  
Adding a small cosmological constant to regulate the long distance
behavior of the metric, we search for all circularly symmetric stationary
vacuum solutions with vanishing angular momentum and approaching the de Sitter metric with a possible 
 angle deficit at infinity. We find a two-parameter family of such geometries. Apart from the cosmological de Sitter horizon, these solutions generally contain
another Killing horizon and should therefore be
interpreted as black holes from the viewpoint of the low-energy 
theory. Contrary to naive expectations, their central singularity is
not resolved by the higher derivative terms present in the action. It is unknown at present if these solutions form as a result of gravitational collapse.
The only
solution regular everywhere is just the de Sitter metric devoid of
any black hole horizon.
\end{abstract}

\maketitle
\section{Introduction}

Despite huge efforts in the last few decades, the formulation of a quantum theory of gravitation remains elusive. In particular, there lingers the open question of how to put together a theory that reproduces the well-known and tested infrared (IR) behavior of general relativity (GR) at the scales of the solar system and cosmology, while having a consistent ultraviolet (UV) limit. While to describe many low-energy systems it is often enough to consider GR as an effective field theory (EFT), where the low-energy Lagrangian is complemented with an infinite series of higher dimensional operators encoding the effect of UV physics, there are situations in which a full description (valid for all ranges of energies) is needed. 

The most prominent of such situations is provided by the existence of singularities within GR. The latter are regions of divergent spacetime curvature, where strong quantum gravitational effects cannot be neglected. The most worrisome singularities are cosmological and those occurring in the interior of black holes (BH). In this paper we focus on the latter. Although there are reasons to believe that BH singularities may always be hidden behind a horizon~\cite{CCC}, thus remaining inaccessible to exterior observers, they are nevertheless the endpoint of any world-line crossing the event   horizon of a BH. Any observer falling into the BH will unavoidably hit the singularity, thus quitting the range of validity of any EFT of gravity.
Resolving the dynamics of the spacetime in the high-curvature region near singularities will therefore require  a theory of quantum gravity.

At present, we still do not have such a theory at our disposal. A possible candidate, which has attracted much interest in  recent years, is quadratic gravity~\cite{Stelle:1977ry,Salvio:2018crh}, 
where the Einstein-Hilbert Lagrangian is complemented by adding terms quadratic in the Riemann tensor, which make it renormalizable \cite{Stelle:1976gc,Fradkin:1981iu,Barvinsky:2017zlx}. However, this theory contains a ghost in the spectrum, as a consequence of the presence of four time derivatives in the action, leading to violation of
unitarity or catastrophic instabilities at high energies. Moreover, BH solutions in quadratic gravity -- such as the  Schwarzschild metric, which is still a solution of the field equations \cite{Lu:2015cqa,Nelson:2010ig}-- are not free of singularities. Therefore, resolution of curvature
singularities within quadratic gravity would require a separate mechanism, unrelated to renormalizability.

A compelling workaround to the ghost problem was proposed in 2009 by Petr Ho\v rava \cite{Horava:2009uw}. If the spacetime is endowed with a preferred time foliation, then one can construct a theory that has only higher order \emph{spatial} derivatives thus avoiding the presence of a ghost. By including in the action all marginal and relevant operators under an anisotropic (Lifshitz) scaling\footnote{Latin indices run over space dimensions only ($i=1,...,d$), while Greek indices include time.}
\begin{align}\label{eq:scaling_ani}
t\rightarrow b^d\ t,\qquad x^i\rightarrow b\ x^i,
\end{align}
where $b$ is constant and $d$ is the number of spatial dimensions, one can make the theory power-counting renormalizable. 

The presence of a time foliation, and thus that of a privileged time direction, make it natural to formulate the theory by using the Arnowitt-Deser-Misner (ADM) decomposition of the metric \cite{PhysRev.116.1322}
\begin{align}\label{eq:ADM}
ds^2=-N^2 dt^2 + \gamma_{ij} (dx^i+N^i dt)(dx^j+N^j dt),
\end{align}
where $N$ and $N^i$ are respectively the lapse function and shift vector,  and $\gamma_{ij}$ is the metric of the spatial hypersurfaces. The presence of a preferred foliation also breaks the full diffeomorphism invariance of GR. The gauge group of the theory reduces to \emph{foliation preserving diffeomorphisms (FDiff)}, i.e. it consists of the direct product of time-dependent spatial diffeomorphisms and time re-parametrizations
\begin{align}
 &t\rightarrow \tilde t(t),\quad x^i\rightarrow \tilde x^i (t,x),
\end{align}
where $\tilde{t}(t)$ is a monotonic function.

Locally, the presence of the preferred foliation breaks Lorentz invariance by allowing  for dispersion relations with higher powers of the spatial momentum, i.e. $\omega^2=c_1^2 k^2+c_2^2 k^4 +... +c^2_{2d}k^{2d}$, with constant coefficients $c_I$. Lorentz invariance may only be recovered as an accidental symmetry in the IR, when higher derivatives are neglected and if the low-energy velocities $c_1$ flow to the same value for all particle species.

The splitting of \emph{FDiff} into two distinct symmetries allows for  formulating two versions of the theory, depending on how one deals with time re-parametrization invariance. In the \emph{projectable} theory, one assumes that the lapse is independent of the spatial coordinates, i.e. a function $N(t)$ of time only. In that case, one can set its value to a constant, which can be chosen to be unity for convenience [$N(t)=1$], gauge fixing  time re-parametrization invariance away. 

After Ho\v rava's seminal paper, there was a surge of activity in understanding the consequences of the theory, and its soundness as a proposal for quantum gravity. Soon, it was realized that the extra propagating scalar mode of the theory -- present alongside the transverse-traceless graviton -- was problematic. Although stable at high energies, in all dimensions higher than $d=2$ it behaves as a tachyon in the IR, signaling that flat space is not a stable vacuum of the theory \cite{Blas:2010hb}. This can be solved by abandoning projectability and allowing the lapse $N(t,x^i)$ to be a function of all spacetime coordinates. This choice
gives rise to a version of the theory known as 
 \emph{non-projectable Ho\v rava gravity}. In this case, new terms are allowed in the Lagrangian, preventing the instability in the IR \cite{Blas:2009qj}. However, it comes at the cost of reintroducing time re-parametrization invariance as a full-fledged gauge symmetry. This causes the presence of an instantaneous propagating mode \cite{Blas:2010hb,Blas:2011ni} and complicates quantization of the model \cite{Donnelly:2011df}, although efforts to pursue this endeavor have not been spared \cite{Orlando:2009en,Giribet:2010th, DOdorico:2014tyh,Bellorin:2019gsc}.

On the other hand, the projectable model has been proven to be fully renormalizable in any spacetime dimension \cite{Barvinsky:2015kil}, preserving gauge invariance to all orders in the loop expansion \cite{Barvinsky:2017zlx}. Moreover, in $2+1$ dimensions it has been shown to be UV-complete \cite{Barvinsky:2017kob}, while propagating a stable non-trivial degree of freedom, thus representing a bona fide theory of     quantum gravity in this dimensionality. There are also some hints that UV-completeness could hold in $3+1$ dimensions \cite{Barvinsky:2019rwn}, although a proof is not yet available.

The existence of BHs has been extensively studied in the IR limit of non-projectable Ho\v rava gravity, where the theory can be shown to be related to Einstein-Aether theory \cite{Jacobson:2004ts,Jacobson:2013xta,vega}. IR BH solutions have a structure similar to those of GR BHs, with a Killing horizon hiding a central singularity where every in-falling world-line ends~\cite{UH1,Blas:2011ni,UH2,UH3,ramos}. Additionally, they also possess a  ``universal horizon''~\cite{UH1,Blas:2011ni}, i.e. a compact
hypersurface of constant preferred time
that surrounds the central singularity
and from which no modes (even instantaneous ones) can escape. Its behavior mimics that of an event horizon in GR \cite{Herrero-Valea:2020fqa,Berglund:2012bu,Berglund:2012fk}. However, as mentioned, these are  \emph{low-energy solutions}, obtained by disregarding  the higher derivative terms that should be important when an observer gets close to the singularity. A complete understanding of the interior of  BHs and of the fate of the universal horizon would require to consider the full Lagrangian of the theory, carefully studying  the effect of the UV-completing terms.

In this work we tie together all these issues, and study the effect of renormalizability on the resolution of curvature singularities  in a controllable playground, that of \emph{projectable Ho\v rava gravity in $2+1$ spacetime dimensions}. As previously mentioned, this is a perturbative UV-complete theory, meaning that \emph{it completely describes gravity, at any energy}.\footnote{Up to, perhaps, presently unknown non-perturbative obstructions.} The Lagrangian functional form of the theory should be valid up to arbitrary high energies -- albeit with varying values for the coupling constants, as predicted by the renormalization group flow. If the solution to BH singularities is linked to the renormalizable character of the theory, without the need for any additional  mechanism, it must then be contained within the dynamics dictated by the action. In other words, we use this model to address the following question: {\it does a UV-complete gravity theory resolve BH singularities?}

In this work we make the first steps in this direction and analyze all circularly symmetric vacuum solutions of the theory with vanishing angular momentum. Recall that GR in (2+1) dimensions does not have any BH solutions with flat or de Sitter asymptotics.\footnote{BHs with anti-de Sitter (AdS) asymptotics do exist in the presence of a negative cosmological constant \cite{Banados:1992wn}. We do not consider this case in the context of projectable Ho\v rava gravity as the constant-lapse condition forces the AdS metric to be time dependent.} By contrast, we find that projectable Ho\v rava gravity admits solutions that can be legitimately called BHs. We start by considering the low-energy limit of the theory and solve the equations of motion analytically when higher derivatives are neglected. The solutions present a Killing horizon and a curvature singularity at the origin of the coordinates. They therefore physically represent BHs, and one can use them as asymptotic solutions far from the center, where the curvature is small and neglecting higher-derivatives is a good approximation. 

We will then attempt to extend our solutions numerically to the higher curvature region of the spacetime, including all the terms in the Lagrangian. By numerical investigation and analytic arguments, we show that there exist no vacuum solutions that are free of central curvature singularities and which reduce far from the center to the BH solutions found in the IR limit of the theory. In other words, we show that the higher-order derivatives, although they make the theory UV-complete and renormalizable, do not resolve the central singularity, at least classically. We will comment on the implications of this finding.

This paper is organized as follows. In Sec.~\ref{sec:projectable_HG} we review 
projectable Ho\v rava gravity. In Sec.~\ref{sec:SphericalSymmetry}, we introduce our 
circularly symmetric and stationary ansatz, which we use to obtain BH solutions in the IR limit of the theory in Sec.~\ref{sec:IR_BH}. The embedding of these BH solutions into the full UV theory is discussed in Sec.~\ref{sec:UV_BH}, where we argue that such UV BH solutions necessarily present a curvature singularity at the center. This is further substantiated in  Sec.~\ref{sec:LambdaExpansion}
by utilizing a boundary-layer expansion.
 We discuss our findings and the mass of our BH solutions in Sec.~\ref{sec:conclusions}. Throughout this paper, we use units in which $\hbar=c=1$, and metric signature $-++$.


\section{projectable Ho\v rava gravity}\label{sec:projectable_HG}
We formulate Ho\v rava gravity in terms of the ADM variables in Eq.~\eqref{eq:ADM}. Under \emph{FDiff} transformations the components of the metric behave as
\begin{align}\label{eq:FDiff}
\nonumber &N\rightarrow N\frac{dt}{dt'},\\
\nonumber &N^i \rightarrow \left(N^j \frac{\partial x'^i}{\partial x^j}-\frac{\partial x'^i}{\partial t}\right)\frac{dt}{dt'},\\
&\gamma_{ij}\rightarrow \gamma_{kl}\frac{\partial x^k}{\partial x'^i}\frac{\partial x^l}{\partial x'^j}.
\end{align}
Their anisotropic scaling dimension under Eq.~\eqref{eq:scaling_ani} is
\begin{align}
[N]=[\gamma_{ij}]=0,\qquad [N^i]=d-1.
\end{align}

The requirement of \emph{FDiff} invariance, time-reversal invariance, parity, power-counting renormalizability under Eq.~\eqref{eq:scaling_ani} 
and  absence of ghosts fixes the action to be
\begin{align}\label{action}
S=\frac{1}{\kappa}\int dt\ d^dx \sqrt{\gamma}\ N \left(K_{ij}K^{ij}-\lambda K^2 -{\cal V}\right),
\end{align}
where $\kappa=16\pi G$ and $\lambda$ are dimensionless coupling constants in the sense of Eq.~\eqref{eq:scaling_ani}
(i.e. they are invariant under that rescaling), and ${\cal V}$ contains 
all possible marginal and relevant operators with respect to the anisotropic scaling. Here, 
$K_{ij}$ is the extrinsic curvature of the slices,
\begin{align}
K_{ij}=\frac{1}{2N}\left(\partial_t \gamma_{ij}-\nabla_i N_j -\nabla_j N_i\right),
\end{align}
with $\nabla_i$ the covariant derivative compatible with $\gamma_{ij}$.

Hereinafter we will focus on the projectable model, thus from now on we will assume that $N(t)$ is independent of the spatial coordinates. We can therefore set it to $N(t)=1$ by exploiting time reparametrization invariance, leaving  time-dependent spatial diffeomorphisms as 
the only remaining gauge symmetry.\footnote{This implies disregarding the global Hamiltonian constraint $\delta S/\delta N=0$.} The potential ${\cal V}$ will  thus be built exclusively out of \emph{FDiff} invariants, 
constructed with the spatial metric and covariant derivatives. Its form in $d=2$ is
\begin{align}
{\cal V}&=2\Lambda + \mu R^2 \,,
\end{align}
where $R$ is the scalar curvature constructed from 
$R_{ijkl}$, the Riemann tensor of the spatial slices. Note that we are omitting a linear term in $R$, 
since it corresponds to the Gauss-Bonnet density in $d=2$, thus being a total derivative. Here $\Lambda$ is the cosmological constant  
which will serve as a regulator for the long-distance behavior of the BH solutions. The total action that we  consider then takes the form 
\begin{align}\label{eq:action}
S=\frac{1}{\kappa}\int dt\ d^2x \sqrt{\gamma}\  \left(K_{ij}K^{ij}-\lambda K^2 -\mu R^2 - 2\Lambda\right).
\end{align}

Although in $2+1$ dimensions GR propagates no local degrees of freedom, this is not the case for Ho\v rava gravity. Due to the reduced symmetry group, 
there is an extra scalar degree of freedom in the spectrum of the theory, with dispersion relation 
\begin{align}
\label{dispflat}
\omega^2=4\mu\ \frac{1-\lambda}{1-2\lambda}\ k^4
\end{align}
around flat-space.
Note the absence of $k^2$ term, as a consequence of the triviality of the piece linear in $R$ in the Lagrangian. Unitarity and stability 
then require $\mu>0$ and either $\lambda<\frac{1}{2}$ or $\lambda> 1$. We will consider the latter case throughout this work, for reasons that will become clear in a moment.

When regarded as a quantum field theory, the action \eqref{eq:action} corresponds to a renormalizable theory. 
Within perturbation theory, all UV divergences can be absorbed by a corresponding redefinition of the coupling constants. 
Correlation functions of observables are then essentially the same as their classical values, with $\kappa, \lambda$ and $\mu$ replaced by $\kappa(k_*),\lambda(k_*),\mu(k_*)$, where $k_*$ is a parameter that sets the interaction scale. The running of the couplings was computed in the one-loop approximation in Ref.~\cite{Barvinsky:2017kob} and reads
\begin{subequations}
\label{eq:betas}
\begin{align}\label{eq:betalambda}
&\frac{d\lambda}{d \log k_*}=\frac{15-14\lambda}{128\pi}\sqrt{\frac{1-2\lambda}{1-\lambda}}\ \tilde{\kappa}\,\\
\label{eq:betacalG} &\frac{d\tilde{\kappa}}{d \log k_*}=-\frac{(16-33\lambda+18 \lambda^2)}{128\pi (1-\lambda)^2}\sqrt{\frac{1-\lambda}{1-2\lambda}}\ {\tilde{\kappa}}^2,
\end{align}
\end{subequations}
where we have defined ${\tilde{\kappa}}=\frac{\kappa}{\sqrt{\mu}}$. 
It can be shown that the flows of $\kappa$ and $\mu$ are separately gauge dependent.
However, those of $\tilde{\kappa}$ and $\lambda$ are independent of the choice of gauge. 
This signals that only these parameters appear in correlation functions of gauge-invariant observables, and in physical observables of the theory. 

The fact that the theory is stable against radiative corrections suggests a justification for attempting a classical treatment down to arbitrarily short scales.
Indeed, the magnitude of quantum fluctuations can be estimated from the action \eqref{eq:action} as follows. Setting that for fluctuations the action is of order unity, $\delta S\sim 1$, and assuming a regular geometry,  we obtain 
\begin{equation}
\label{deltaKR}
(\delta K_{ij})^2\sim \frac{\kappa}{\tau l^2}~,~~~~
(\delta R)^2\sim \frac{\kappa}{\mu\tau l^2}\;,
\end{equation}
where $\tau$ and $l$ are the characteristic time and length scales of the perturbations. The fluctuations of the extrinsic and intrinsic curvatures are related to the metric fluctuations as $\delta K_{ij}\sim \tau^{-1}\delta\gamma_{ij}$, $\delta R\sim l^{-2}\delta\gamma_{ij}$. Substituting into \eqref{deltaKR} and taking the product to get rid of $\tau$ and $l$, we obtain
\begin{equation}
\delta\gamma_{ij}\sim \sqrt{\tilde\kappa}\;,
\end{equation}
which remains small at all scales, as long as $\tilde\kappa$ is small.

Note that the set of $\beta$-functions \eqref{eq:betas} contains a fixed point of the renormalization group flow 
in the region $\lambda>1$ for the values 
\begin{align}\label{fixed_point}
    \lambda_{\bullet}=\frac{15}{14},\qquad \tilde{\kappa}_{\bullet}=0.
\end{align}
This shows that the theory enjoys asymptotic freedom at high energies, thus representing a perturbatively UV-complete quantum field theory. A second fixed point appears at $\lambda=1/2$. 
However, in the vicinity of that point the 
expansion parameter is $\tilde{\kappa}(1-2\lambda)^{-\frac{1}{2}}$ and it remains arbitrary at one loop.
Thus, it cannot be said whether this fixed point persists or not unless a two-loop computation is performed. We will therefore focus on the first fixed point.

After fixing the lapse to $N(t)=1$ by using the projectability condition, the dynamical variables left in the theory are the shift $N^i$ 
and the spatial metric $\gamma_{ij}$. Varying the action with respect to them, we get the following equations of motion
\begin{subequations}
\label{eq:eom}
\begin{align}
{\cal P}_i&\equiv\nabla^j K_{ij}-\lambda \nabla_i K=0,\\
{\cal G}^{ij}&\equiv\nonumber -D_t\left( K^{ij}-\lambda \gamma^{ij} K  \right)-(1-2\lambda) K K^{ij}-2K^{ik}K_k^j\\
\nonumber &+\frac{1}{2} K^{kl}K_{kl}\gamma^{ij}+\frac{\lambda}{2} K^2\gamma^{ij}+\frac{\mu}{2} R^2 \gamma^{ij}+2\mu \Delta R \gamma^{ij}\\
&-2\mu \nabla^i \nabla^j R 
 - \Lambda \gamma^{ij} =0\;,
\end{align}
\end{subequations}
where we have used that $R^{ij}=R\gamma^{ij}/2$ in two dimensions. Here the covariant time derivative is defined as 
\be
\label{Lieder}
D_t=\partial_t-{\cal L}_{\vec N}\;, 
\ee
where ${\cal L}_{\vec N}$ is the Lie derivative along the shift vector, so that for a two-index tensor we have,
\be
D_t A^{ij}=\partial_t A^{ij}-N^k\nabla_k A^{ij} 
+A^{ik}\nabla_kN^j+A^{jk}\nabla_kN^i\;.
\ee
On top of this and like in any gauge theory, local invariance under spatial time-dependent diffeomorphisms \eqref{eq:FDiff} imposes 
a ``Bianchi'' identity~\cite{bianchi_jacobson,UH1,ramos}
\begin{equation}\label{eq:Bianchi}
\nabla_j {\cal G}^{ji}+\gamma^{ij}D_t {\cal P}_j
+K {\cal P}^i=0\;.
\end{equation}

We assume the cosmological constant to be small compared to the UV scale set by $\mu^{-1}$, $\Lambda\mu\ll 1$. It is needed to regulate the long-distance behavior of the solutions. This is a peculiarity of (2+1) dimensions, where the gravitational field of a localized source does not vanish at infinity even in GR, persisting as a global angle deficit. We find that the problem gets aggravated in Ho\v rava gravity, where in the limit 
$\Lambda\rightarrow 0$ the conical deficit grows indefinitely at large radii, despite the fact that all curvature invariants tend to zero. Introduction of non-vanishing $\Lambda$ turns this into a well-behaved de Sitter asymptotics with a finite angle deficit.


\section{\label{sec:SphericalSymmetry}Circularly symmetric spacetimes}

In the following, we will write the equations of motion for a general non-rotating circularly symmetric and stationary ansatz. We use polar coordinates $(r,\theta)$ for the spatial slices and write the ADM metric in the preferred foliation as
\begin{align}
    ds^2=(-1+N_iN^i)dt^2 +2 N_idx^i dt +\gamma_{ij}dx^i dx^j\,,
\end{align}
where we have already fixed $N(t)=1$.

Stationarity of the solution imposes $\partial_t N^i=\partial_t \gamma_{ij}=0$, while the requirement of circular symmetry enforces $N^\theta=0$. Finally, any two-dimensional metric is conformally flat, implying that $\gamma_{ij}$ can only depend on a single function $G(r)$. We thus adopt, without loss of generality, the ansatz
\begin{align}
\label{eq:metransatz}
    ds_{2}^2=dr^2+r^2 G(r)^2 d\theta^2
\end{align}
for the two-dimensional spatial metric.

Putting all this together and defining $N^r=F(r)$, our ansatz for the full metric finally takes the form
\begin{align}\label{eq:ansatz}
    ds^2=(-1+F(r)^2)dt^2 +2 F(r) dt dr +dr^2+r^2 G(r)^2 d\theta^2.
\end{align}
This chart of coordinates is reminiscent of the well-known Gullstrand-Painlev\'e coordinates (see e.g. Ref.~\cite{Nielsen:2005af}) in standard solutions -- e.g. the Schwarzschild metric and the  Banados, Zanelli and Teitelboim (BTZ) BHs \cite{Banados:1992wn}.

We now insert this ansatz into the equations of motion \eqref{eq:eom}. From \(\mathcal{P}_r\) and \(\mathcal{G}_{\theta\theta}\) we obtain differential equations that are second order in derivatives for \(F(r)\), and fourth order for \(G(r)\). Since the precise form of the equations is cumbersome and not very illuminating, we relegate them to Appendix \ref{sec:ExplicitEOMs}. Schematically, their form is
\begin{subequations}
\label{eq:ODEs}
\begin{align}
&E_1 \left[ F, F', F'', G, G', G'' \right] =0,
\label{eq:Feq}
\\
&E_2 \left[ F, F', F'', G, G', G'', G^{(3)}, G^{(4)} \right] =0,
\label{eq:Geq}
\end{align}
\end{subequations}
where a prime denotes a derivative with respect to the argument of the function. From now on we will suppress the arguments for clarity whenever needed.

From \(\mathcal{G}_{rr}\) we can in principle obtain another second order equation for $F(r)$. However, one can combine it with Eq.~\eqref{eq:Feq} to eliminate $F''(r)$ and rewrite it as a constraint
\begin{equation}
E_3 \left[ F, F', G, G', G'', G^{(3)} \right] =0.
    \label{eq:Constraint}
\end{equation}
Furthermore, using the Bianchi identity \eqref{eq:Bianchi}, one can show that
\begin{equation}
    E'_3 +2 G r F' E_1 +\left(\frac{G'}{G}+\frac{1}{r}\right) (E_2-2E_3)=0.
    \label{eq:ConstraintPropagation}
\end{equation}

Therefore, we see that the system is {\it not} over-determined. The condition \eqref{eq:ConstraintPropagation} implies that once the constraint equation \eqref{eq:Constraint} is imposed at a point, e.g. at a boundary \(r_0\), then the constraint is propagated throughout \(r\), provided that the equations of motion \eqref{eq:ODEs} are satisfied.

Close examination of Eqs.~\eqref{eq:ODEs} and \eqref{eq:Constraint} reveals that they are invariant under constant rescalings of $G(r)$
(i.e., if $(F,G)$ is a solution, also 
  $(F,k G)$, with $k$ a constant rescaling factor,
  is a solution to the same theory). Thus, we can take advantage of this and further simplify the equations of motion by defining a new variable
\begin{align}\label{eq:GammaDef}
\Gamma (r) = \dfrac{1}{r}+\dfrac{G'(r)}{G(r)},    
\end{align}
This reduces Eq.~\eqref{eq:ODEs} to a third-order system in \(\Gamma(r)\) of the form
\begin{subequations}\label{eq:GammaODEs}
\begin{align}
\label{eq:GammaODEs1}
    \mathcal{E}_1 & \equiv (\lambda -1) \left(F''+F'\Gamma  +F \Gamma
   '\right)+F \Gamma '+F\Gamma ^2    
   = 0,\\
\label{eq:GammaODEs2}
    \mathcal{E}_2 & \equiv \mu  \left(-8\Gamma'''-16 \Gamma  \Gamma ''-12 \left(\Gamma'\right)^2
    +8 \Gamma ^2 \Gamma '+4 \Gamma ^4\right)\nonumber\\
   +&(\lambda -1) \left(2 F F''+\left(F'\right)^2
   +4 F F' \Gamma + 2 F^2 \Gamma '
   +F^2 \Gamma ^2 \right)\nonumber\\
   +&2 F F'' +2\left(F'\right)^2
   -2\Lambda
   = 0,
\end{align}
while the constraint \eqref{eq:Constraint} becomes a second-order equation in \(\Gamma(r)\),
\begin{align}\label{eq:GammaConstraint}
    \mathcal{E}_3  \equiv & \mu  \left(8 \Gamma  \Gamma ''
    -4 \left(\Gamma'\right)^2
    +8 \Gamma ^2 \Gamma '-4 \Gamma
   ^4\right)\nonumber\\
   &+(\lambda -1) \left(-\left(F'\right)^2
   -2 F F'\Gamma-F^2\Gamma ^2
   \right)\nonumber\\
   &-2 F F' \Gamma 
   + 2 \Lambda
   = 0.
\end{align}
\end{subequations}
Henceforth, instead of dealing with the original equations $\mathcal{E}_1$ and $\mathcal{E}_2$, we can instead solve the system consisting of \(\mathcal{E}_1\) and \(\mathcal{E}_3\)
(i.e. the system consisting of one of the evolution equation and the constraint equation), effectively dealing with a \emph{second-order} system in both \(F(r)\) and \(\Gamma (r) \), and thus requiring only four integration constants.
One can always do this because the Bianchi identity ensures that the remaining equation \(\mathcal{E}_2\) will be satisfied by the solution. 

In the case of vanishing cosmological constant $\Lambda$, we can identify two symmetries of the equations of motion (besides the aforementioned invariance under constant rescalings of $G$), corresponding to shifts and rescaling of the radial coordinate. In more detail, the  field equations are invariant under
\begin{subequations}
\label{eq:scalings}
\begin{align}\label{eq:scaling}
    &F(r)\mapsto b F(b\,r +a),\\
    &\Gamma(r)\mapsto b\Gamma(b\,r+a),\label{eq:scaling2}
\end{align}
\end{subequations}
with arbitrary constants $a$ and $b$. This symmetry will play an important role in the numerical analysis of Sec.~\ref{sec:UV_BH}.

\subsection{\label{sec:BlackHoles} Black Holes}

We will define the concept of a BH  from the perspective of an observer in the IR limit of the theory, in analogy to the general relativistic case. If we were dealing with GR, then a BH would be characterized by the presence of a trapped surface for null trajectories~\cite{penrose}, i.e. for (massless) particles with dispersion relation $\omega= k$.
The outermost of all trapped surfaces is usually referred to as the apparent horizon, which in stationary circularly symmetric configurations
coincides with the Killing and event horizons.
By analogy, we will assume that in the IR limit of Ho\v rava gravity, massless particles move with dispersion relation $\omega=k+{\cal O}(k^2)$, and therefore we will borrow the same definition of a BH. 

The Killing horizon can be identified by requiring the time-like Killing vector  ${\partial}/{\partial t}$
to have vanishing norm at the position of the horizon. For our metric ansatz, this leads to the condition
\begin{equation}\label{eq:HorizonCondition}
\frac{\partial}{\partial t}\cdot \frac{\partial}{\partial t}=g_{tt}=  -1+F^{2}(r)=0\,.
\end{equation}
Since our ansatz for the metric is stationary, the Killing horizon coincides with the apparent horizon -- which is defined in turn by the
constant \(r\) surfaces becoming null \(g^{\mu\nu}\partial_\mu r \partial_\nu r=0\)~\cite{Nielsen:2005af} -- and with the event horizon.

It is worth  noting here an important difference with known BH solutions in \emph{non-projectable} Ho\v rava gravity \cite{UH1,Blas:2011ni}. In that theory, solutions are characterized by the presence of a \emph{universal horizon}, a compact surface that traps all signals, regardless of their dispersion relation. Its position can be identified by requiring the unit-vector orthogonal to the foliation,
\begin{align}
    U_\m =-N \delta_{\m}^t,
\end{align}
to become orthogonal to the Killing vector $\partial_t$ (which is tangent to
hypersurfaces of constant $r$), i.e.
\begin{align}
U\cdot \frac{\partial}{\partial t}=-N=0.
\end{align}
Because of the condition $N(t)=1$, it is impossible for solutions in projectable Ho\v rava gravity to present universal horizons. Thus, we can expect signals of arbitrary speed to be able to eventually probe the interior of the BH (as defined in the IR) and escape from it.

\section{Black Holes in the IR limit}
\label{sec:IR_BH}
We will  now face the issue of obtaining circularly symmetric solutions to the equations of motion \eqref{eq:GammaODEs}--\eqref{eq:GammaConstraint}. This is not an easy task in general. The non-linear character of the equations renders the problem hard to tackle analytically. However, there is a regime in which 
solutions can be found rather easily, namely the IR limit of the theory, which one can obtain by setting $\mu=0$. Solutions obtained in this way will be valid whenever the spatial curvature of the slices is low. This corresponds to focusing on the region $r\gg \sqrt{\mu}$, where we expect this to happen and where higher derivative terms can be ignored. 

We start by considering the combination $({\cal E}_2+{\cal  E}_3)/2-F{\cal E}_1$ of the equations, which yields
\begin{align}\label{eq:IReq1}
    FF''+(F')^2-FF'\Gamma-F^2\Gamma'-F^2\Gamma^2=0 .
\end{align}
Assuming that $F(r)$ is non-vanishing everywhere and introducing a new variable
\begin{equation}\label{Ydef}
    Y=\Gamma-\frac{F'}{F} \,,
\end{equation}
this equation can be cast into the simple form
\begin{equation}\label{Yeq}
    -Y'+2Y^2-3Y\Gamma=0    .
\end{equation}
Note that the derivative of $\Gamma$ has disappeared from the equation.

Let us first consider the solution $Y=0$ to this equation, which implies
$\Gamma=F'/F$. Substituting this relation in Eq.~\eqref{eq:GammaConstraint} (with $\mu=0$),
we find that
the function $F$ is linear,
\begin{equation}\label{FdeSit}
    F=\pm r\sqrt{\frac{\Lambda}{2\lambda-1}}  .
\end{equation}
Note that this solution exists for positive $\Lambda$ only if
$\lambda>1/2$, and that it corresponds to a constant radial function
\begin{equation}\label{GdeSit}
    G(r)= G_{\infty}.
\end{equation}
For $G_\infty=1$, this reduces to the de Sitter metric, which is regular everywhere and has a cosmological Killing horizon at 
\begin{align}\label{rds}
    r_{\rm dS}=\sqrt{\frac{2\lambda-1}{\Lambda}}
\end{align}
Other choices of $G_{\infty}$ lead to an angle deficit and  a conical singularity at the origin. 

We now consider
the case of non-vanishing solutions to Eq.~\eqref{Yeq}, $Y\neq0$.
We assume $Y>0$
without loss of generality.\footnote{Positive $Y$
can always be achieved by changing the sign of $r$, as manifest from the definitions \eqref{Ydef} and \eqref{eq:GammaDef}.} 
 From Eqs.~\eqref{Ydef} and \eqref{Yeq}, we can express
$\Gamma$ and $F'/F$ in terms of $Y$ and its derivative,
\begin{align}\label{GYFY}
\Gamma&=-\frac{Y'}{3Y}+\frac{2}{3}Y, & \frac{F'}{F}&=-\frac{Y'}{3Y}-\frac{Y}{3} .
\end{align}
The second of these equations can be integrated if we introduce a new
function $X(r)$ such that 
\begin{align}\label{XY}
    X'=Y  \,,
\end{align}
yielding
\begin{align}\label{FXY}
    F=\frac{C}{Y^{1/3}}\,e^{-X/3} \, ,
\end{align}
where $C$ is an integration constant. Substituting this and the first relation
\eqref{GYFY} into Eq.~\eqref{eq:GammaODEs}, we obtain a differential equation
involving $X$ and $Y$,
\begin{align}\label{E3deSit}
    0=&-\frac{(4\lambda-2)(Y')^2}{9Y^2}+\frac{(4\lambda-2)Y'}{9}
    +\frac{(5-\lambda)Y^2}{9}\nonumber\\
    &+\frac{2\Lambda}{C^2}\, Y^{2/3}\,e^{2X/3}  .
\end{align}
Recalling that $Y$ is the derivative of $X$, we observe that this is a
second-order differential equation for the function $X(r)$. 
Importantly, this equation does not contain explicitly the variable
$r$, and thus can be reduced to a first-order equation if we choose
$X$ to be our independent variable, instead of $r$. We therefore substitute
\begin{align}
    Y'=\frac{dY}{dX}\,X'=\frac{dY}{dX}\,Y  
\end{align}
and obtain
\begin{align}\label{YXeq}
  \nonumber  0=&-\frac{2(2\lambda-1)}{9}\bigg(\frac{dY}{dX}\bigg)^2
    +\frac{2(2\lambda-1)}{9} Y\frac{dY}{dX}+\frac{5-\lambda}{9}Y^2
    \\
    &+\frac{2\Lambda}{C^2}Y^{2/3}e^{2X/3}.
\end{align}
This is further simplified by the definition $Y={\hat  Y}^{3/2}e^{X/2}$, which yields
\begin{align}\label{hatYeq}
    -(2\lambda-1)\bigg(\frac{d\hat Y}{dX}\bigg)^2+\hat Y^2+\frac{4\Lambda}{C^2}=0  .
\end{align}
Again,  in the case of positive $\Lambda$ (on which we focus in this paper) the solution exists only
if $\lambda>1/2$. Solving for $\hat Y$ and substituting into the
expression for $Y$ we find
\begin{align}\label{Ysol}
    Y=\bigg(\pm\frac{2\sqrt\Lambda}{|C|}\sinh\frac{X-X_0}{\sqrt{2\lambda-1}}\bigg)^{3/2}e^{X/2}\,,
\end{align}
where $X_0$ is an integration constant, and the signs \(\pm\) are chosen to have the expression in brackets positive. The constant \(X_0\) can be absorbed in the shift of $X$ and subsequent rescaling $C\mapsto Ce^{X_0/3}$, which leave both $Y$ and $F$ invariant. 
Therefore we set $X_0=0$ henceforward. 

The solutions presents two
branches corresponding  to the plus/minus sign and positive/negative $X$. 
Let us focus
on the case $X<0$ (we will comment on the branch with $X>0$ at the end of the section). Using Eqs.~\eqref{XY}, \eqref{FXY} and the first
of Eqs.~\eqref{GYFY}, we obtain the solution in parametric form
\begin{subequations}\label{solIR}
\begin{align}
\label{solIRr}
&\sqrt{\Lambda}\, r=\frac{B}{2}\int_{-\infty}^{X}
\frac{e^{-X'/2}}{\Big(\sinh{\frac{-X'}{\sqrt{2\lambda-1}}}\Big)^{3/2}}dX',\\
\label{solIRF}
&F=\pm B
\frac{e^{-X/2}}{\Big(\sinh{\frac{-X}{\sqrt{2\lambda-1}}}\Big)^{1/2}},\\
\label{solIRG}
&rG=G_\infty B\sqrt{\frac{2\lambda-1}{\Lambda}} 
\frac{e^{X/2}}{\Big(\sinh{\frac{-X}{\sqrt{2\lambda-1}}}\Big)^{1/2}},
\end{align}
\end{subequations}
where we have introduced a new integration constant $G_\infty$ and 
defined
\begin{equation}\label{BC}
    B=\bigg(\frac{|C|^3}{2\sqrt\Lambda}\bigg)^{1/2}.
\end{equation}

The $\pm$  in Eq.~\eqref{solIRF} corresponds to the sign of the original integration constant
$C$, which can be both positive or negative, whereas $B$ is strictly
positive. Note also that $B$ is dimensionless and that the integral in
Eq.~\eqref{solIRr} converges at the lower end, as long as $\lambda<5$,
and diverges as $X\to 0^{-}$, 
so that $r$ varies from $0$ to $+\infty$. This is a relevant range for $\lambda$, as it includes the fixed point \eqref{fixed_point}, and we will focus on it in the following.

Let us study the asymptotics of the solution \eqref{solIR}. Consider
first $X\to0^{-}$, corresponding to $r\to +\infty$, which yields
\begin{subequations}\label{solITlarger}
\begin{align}
    &\sqrt\Lambda\,r\approx (2\lambda-1)^{3/4}\frac{B}{\sqrt{-X}},\\
&F\approx \pm (2\lambda-1)^{1/4}
\frac{B}{\sqrt{-X}}\approx\pm\sqrt{\frac{\Lambda}{2\lambda-1}}\,r,\\
&rG \approx
G_\infty\frac{(2\lambda-1)^{3/4}}{\sqrt{\Lambda}}\frac{B}{\sqrt{-X}}
\approx G_{\infty}r.
\end{align}
\end{subequations}
In these expressions we recognize the de Sitter metric [Eqs.~\eqref{FdeSit} and
\eqref{GdeSit}] with the deficit angle set by $G_\infty$. The
integration constant $B$ has dropped out. Therefore, at sufficiently large radii, the solution given by Eq.~\eqref{solIR} approaches  the de Sitter geometry.

Second, we consider the other extreme $X\to -\infty$, corresponding to
$r\ll B/\sqrt{\Lambda}$. In this limit, we obtain
\begin{subequations}\label{solIRsmallr}
\begin{align}
\label{rIRsmallr}
&\sqrt\Lambda\,
r\approx\frac{2\sqrt{2\lambda-1}}{3-\sqrt{2\lambda-1}}\sqrt{2}B
\exp\bigg[\bigg(\frac{3}{2\sqrt{2\lambda-1}}-\frac{1}{2}\bigg)X\bigg],\\
\label{FIRsmallr}
&F\approx\pm\sqrt{2}B
\exp\bigg[\bigg(\frac{1}{2\sqrt{2\lambda-1}}-\frac{1}{2}\bigg)X\bigg]= \pm F_0(\sqrt{\Lambda}\,r)^{-\sigma},\\
\label{GIRsmallr}
&G\approx \frac{G_\infty B}{r}\sqrt{\frac{2(2\lambda-1)}{\Lambda}}
\exp\bigg[\bigg(\frac{1}{2\sqrt{2\lambda-1}}+\frac{1}{2}\bigg)X\bigg]\nonumber \\
&\quad = G_0 (\sqrt{\Lambda}\,r)^{2\sigma},
\end{align}
\end{subequations}
where
\begin{align}\label{sigma}
&\sigma=\frac{\lambda-2+\sqrt{2\lambda-1}}{5-\lambda}>0\;,\\
\label{F0B}
&F_0=(\sqrt{2}B)^{1+\sigma}
\bigg(\frac{3-\sqrt{2\lambda-1}}{2\sqrt{2\lambda-1}}\bigg)^{-\sigma}\;,\\
\label{G0B}
&G_0=G_\infty\sqrt{2\lambda-1}\, (\sqrt{2}B)^{-2\sigma}
\bigg(\frac{3-\sqrt{2\lambda-1}}{2\sqrt{2\lambda-1}}\bigg)^{1+2\sigma}\;.    
\end{align}

Remarkably, these solutions present a second Killing horizon besides the cosmological one, located at 
\begin{align}
    r_H=\frac{F_0^{\frac{1}{\sigma}}}{\sqrt{\Lambda}}.
\end{align}
From the point of view of the low-energy theory, this solution then describes a black (white) hole for positive (negative) $F(r)$. The second horizon is well within the de Sitter radius \eqref{rds} as long as $F_0\ll 1$, which holds if $B\ll 1$. Note that $B\ll 1$ also implies $G_0\gg G_\infty$.

Were these solutions to be trusted in the whole spacetime, extrapolating them inwards would lead to curvature singularities at the origin, as can be seen in the different curvature scalars
\begin{align}
\label{eq:extrK}
&K=-\left(\frac{F_0}{\Lambda^{\sigma/2}}\right)\frac{(1+\sigma)}{r^{1+\sigma}},\\
&R=-\frac{4\sigma(1+2\sigma)}{r^2}, \label{eq:spatial_curv}\\
&K_{ij}K^{ij}=\left(\frac{F_0^2}{\Lambda^{\sigma}}\right)\frac{(5\sigma^2+4\sigma+1)}{r^{2+2\sigma}},
\end{align}
as well as in the full three-dimensional spacetime curvature
\begin{align}\label{eq:Riccie3}
    R^{(3)}&=2\sigma(1+2\sigma)\bigg[\left(\frac{F_0^2}{\Lambda^{\sigma}}\right)\frac{1}{r^{2+2\sigma}}-\frac{2}{r^2}\bigg],
\end{align}
where we have made use of the Gauss--Codazzi relations.

In light of all this, we refer to these solutions as ``IR BHs''. We can think of them as akin to the Schwarzschild--de Sitter BHs of GR. They are solutions to the low energy limit of a gravitational theory,
and behind a horizon they include a region (near the center) where the description provided by the low-energy Lagrangian breaks down, and  where one  therefore needs to account for the dynamics of the full theory. In particular, in the case at hand we expect the effects of the $\mu R^2$ term to become important at a distance $r\sim \sqrt{\mu}$ from the center. This ensures that for any IR solution there is always a value of $\sqrt{\mu}\ll r_H$ for which the UV corrections are only important deep inside the geometry. Therefore, they do not modify the horizon, and the solution  still appears as a BH to exterior observers.

Let us discuss the limit of 
vanishing cosmological constant $\Lambda\rightarrow 0$.
If one keeps the combinations 
$B\Lambda^{-\frac{\sigma}{2(1+\sigma)}}$, 
$G_\infty \Lambda^{-\frac{\sigma^2}{(1+\sigma)}}$
fixed in this limit, the BH horizon radius $r_H$ remains finite, whereas the de Sitter asymptotics
are pushed to infinity. The solution \eqref{FIRsmallr}-\eqref{GIRsmallr} is then valid for arbitrary large radii.
Note that the  curvature invariants
$K$, \(K_{ij}K^{ij}\), \(R\) and \(R^{(3)}\) of this solution vanish as
$r\to\infty$, just like the projections of
the Riemann tensor on the (normalized) timelike and spacelike Killing vectors (parallel respectively to $\partial_t$ and $\partial_\theta$). This implies that the asymptotic geometry is locally flat, but not globally such, because it presents an asymptotically increasing negative deficit angle due to the growth of the function $G$.
 This unappealing behavior is due to the peculiarity of the 2-dimensional spatial geometry, where a change of the circumference of a circle does not affect the local characteristics of space. We have seen how this long-distance behavior is regulated by the presence of a positive cosmological constant. 

Finally, we comment on the $X>0$ branch of solutions. 
A similar analysis shows that it also presents de Sitter asymptotics for \(X \rightarrow 0^+\). However, the solutions in this branch do not have any additional Killing horizons, besides the cosmological one. Moreover, the metric function $rG$ diverges at $r\rightarrow 0$, a behavior that
appears rather pathological. For these reasons, we are not going to consider these solutions further in this
paper.

\section{\label{sec:UV_BH} Black Holes in the UV-complete theory}

We now analyze how the inclusion of the higher derivative terms affects the BH solutions found in the previous section. The higher derivative terms are important at distances $r\sim \sqrt\mu\ll 1/\sqrt\Lambda$. Therefore, in this section we neglect the cosmological constant and use Eqs.~\eqref{FIRsmallr}, \eqref{GIRsmallr} as the large-distance form of the solution. In other words, we will look for solutions of Eqs.~\eqref{eq:GammaODEs1}, \eqref{eq:GammaConstraint} with $\Lambda=0$, which have asymptotics 
\begin{subequations}
\label{eq:asymptotics}
\begin{align}
    &F(r)= F_\infty r^{-\sigma},\\
    &\Gamma(r)=\frac{1+2\sigma}{r},
    \label{eq:asymptoticsGamma}
\end{align}
\end{subequations}
at $r \rightarrow \infty$, where $F_\infty$ is a constant. To simplify notations, we will from now on measure distances in units of $\sqrt{\mu}$, which corresponds to formally setting $\mu=1$ in the equations.

Based on generic arguments, one may expect 
 a renormalizable UV-complete theory including higher order spatial derivatives to allow for ``resolving'' the central  curvature singularity of BHs, which is also present in the aforementioned IR solutions (see e.g. Ref.~\cite{Blas:2014aca} for a mention of this possibility).
In the following, we will therefore try to seek BH solutions to the UV-complete theory that implement this feature.

In principle, the singularity could be resolved in (at least) three possible ways: {\it (a)} the full solution may have a regular center at $r=0$; {\it (b)} the full solution may extend all the way down to $ r=-\infty$, where it may open up into another asymptotic region, thus describing a wormhole configuration; {\it (c)} the full solution extends all the way down to $  r=-\infty$, with the metric function $rG$ and all curvature invariants remaining bounded: this would describe the resolution of the singularity into an infinite throat. 

To see if any of these options gets realized, we start by counting the number of free parameters in the solution of the system \eqref{eq:GammaODEs1}, \eqref{eq:GammaConstraint} once the large-distance asymptotics \eqref{eq:asymptotics} are fixed. We linearize the functions $F$, $\Gamma$ around their asymptotics by writing
\begin{subequations}\label{asexpand}
\begin{align}
    F(r)&=F_\infty r^{-\sigma}\big(1+f( r)\big), \\
    \Gamma(r)&=\frac{1+2\sigma}{ r}\big(1+g(r)\big)\,,
\end{align}
\end{subequations}
where we assume $f$ and $g$ to be small at large $r$. Substituting this into Eqs.~\eqref{eq:GammaODEs1}, \eqref{eq:GammaConstraint} and expanding to linear order in $f$ and $g$ we find a system of two second-order linear equations, whose general solution reads (see Appendix~\ref{sec:WKB} for details),
\begin{subequations}\label{fggen}
\begin{align}
f&=f_\infty  r^{-2(1-\sigma)}+C_1\frac{\sigma}{r}+C_2+C_3 f_3(r)+C_4 f_4(r),\\
g&=g_\infty r^{-2(1-\sigma)}+C_1\frac{1}{r}+C_3 g_3(r)+C_4 g_4(r)\;.
\end{align}
\end{subequations}
Here the coefficients $f_\infty$, $g_\infty$ are fixed in terms of $\lambda$ and $F_\infty$ and correspond to a particular solution of the linear system. Notice that
consistency of the asymptotic expansion requires that these solutions decrease at $r\to \infty$, which implies $\sigma<1$. This requirement is satisfied if $\lambda<5/2$, which includes the interesting fixed point \eqref{fixed_point}.

The coefficients $C_I$ in Eqs.~\eqref{fggen} are arbitrary. The parameters $C_1$, $C_2$ correspond to the symmetry \eqref{eq:scalings} of the equations, spontaneously broken by the asymptotic form \eqref{asexpand}. One easily recognizes in the linearly independent solutions they multiply the results of an infinitesimal shift and rescaling of the asymptotics \eqref{asexpand}. 
The two remaining linearly independent solutions $(f_3,g_3)$ and $(f_4,g_4)$ are oscillating and can be found analytically in the limit $(\lambda-1)\ll 1$ using a version of the WKB expansion \cite{Bender:1977} (see Appendix \ref{sec:WKB}). Importantly, the amplitude of $g_3$ and $g_4$ grows at large $r$,
destroying the desired asymptotic behavior. To satisfy the boundary conditions at infinity, we have to set $C_3=C_4=0$. Thus, we conclude that imposing the large-distance asymptotics \eqref{asexpand} leaves only 2 free parameters $C_1$, $C_2$, both corresponding to the exact symmetries of the equations.

As the next step, we include non-linear corrections to the asymptotic expansion. Motivated by the results of our linearized analysis, we use an ansatz for $F$ and $\Gamma$ in the form of a double series in inverse powers of $r$,
\begin{subequations}\label{eq:IR_series}
\begin{align}
    &F(r) = F_\infty r^{-\sigma}  \left[1 + \sum_{n,m } \dfrac{f_{(n,m)}}{r^{n + m\sigma}}\right],\\
    &\Gamma (r) = \dfrac{1+2\sigma}{r} \left[ 1 + \sum_{n,m}^{\infty}  \dfrac{g_{(n,m)}}{r^{n +m\sigma}}\right],
    \label{eq:IR_seriesGamma}
\end{align}
\end{subequations}
with \(n\), and \(m\) integers such that \(n+m\sigma > 0\). Plugging this ansatz into the equations of motion, the latter can be solved perturbatively in powers of $r^{-1}$, in terms of only two integration constants $F_\infty$ and $f_{(1,0)}$. 

\begin{figure}
\includegraphics[width=0.5\textwidth]{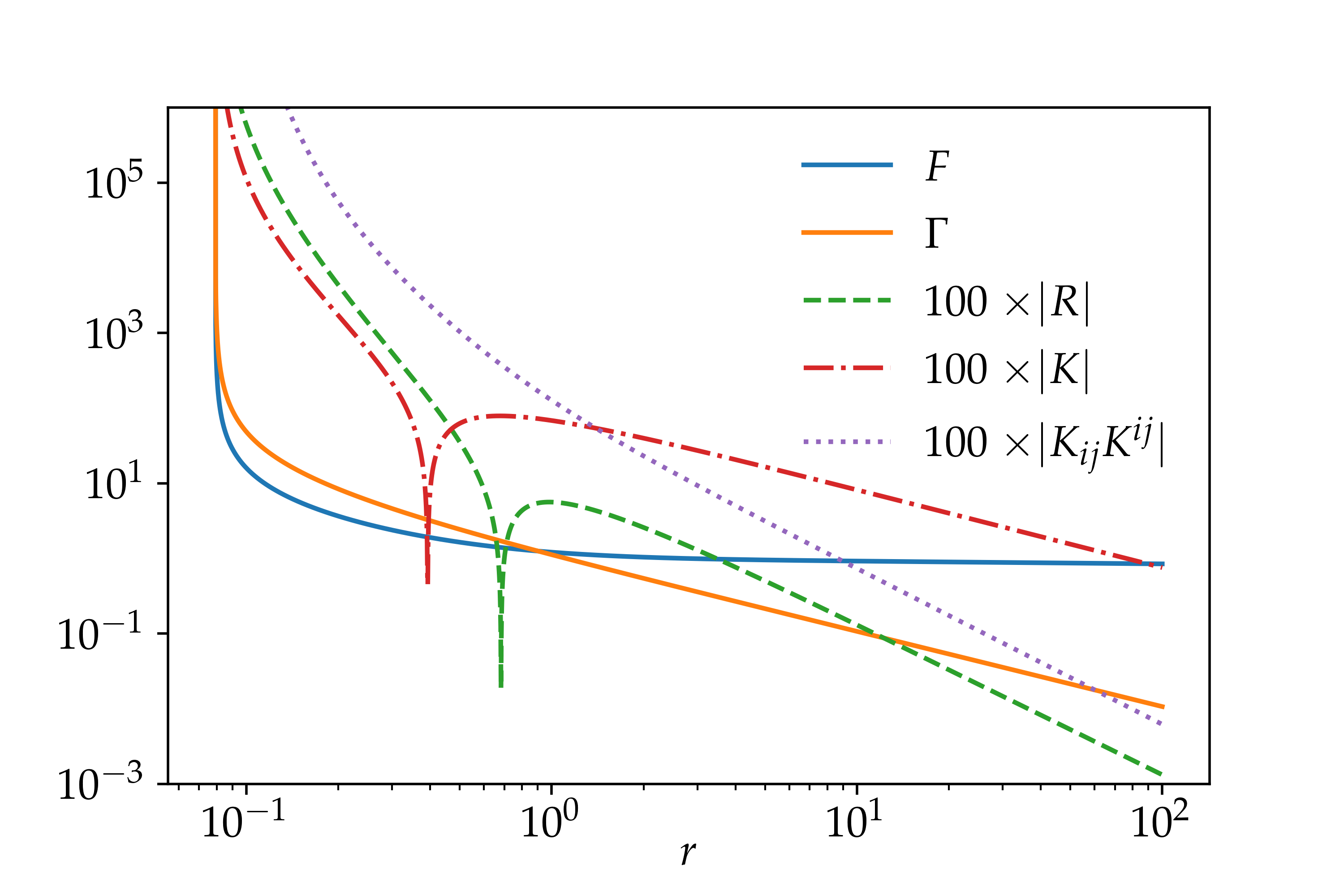}
\caption{Metric functions integrated inwards and the corresponding curvature invariants for \(F_{\infty} = 1\), \(f_{(1,0)} = 0\) and \(\lambda = \frac{15}{14}\). The qualitative behavior of the solution is the same for other values of $\lambda$.}
\label{fig:FGamma}
\end{figure}

We now fix $F_\infty=1$, $f_{(1,0)}=0$ and numerically integrate Eqs.~\eqref{eq:GammaODEs1}, \eqref{eq:GammaConstraint} from large $r$ towards the center. The result is shown in Fig.~\ref{fig:FGamma}. We see that $F$
and $\Gamma$ monotonically grow as $r$ decreases and diverge at a finite value of $r$. (Note however that the areal radius $|r G|$ goes to zero as 
 $F$ and $\Gamma$ diverge, i.e. the area of the singularity vanishes.) The curvature invariants also diverge at that point, indicating that the BH singularity persists even after the inclusion of the higher derivative terms. Notice that varying $F_\infty$ and $f_{(1,0)}$ will not change this result, aside from rescalings/shifts of the solutions. As discussed above, these parameters correspond to exact symmetries that cannot turn a singular behavior into a regular one.

However, it is still premature at this point to claim the absence of a regular solution with certainty because of the following caveat in the above argument. It is logically possible that the divergence observed in the numerical solution is due to a spurious admixture (produced by numerical errors) of modes that are regular throughout the spacetime
and modes that are instead singular. 
Thus, we need to further scrutinize our numerical procedure to ensure its stability.

Let us focus on the possibility that the BH might have a regular center (option {\it (a)} above).
We thus impose regularity by assuming that
$ F(r)$ and $ G(r) $ are analytic near $r=0$,
with respectively only odd and even powers of $r$~\cite{Alcubierre:1138167}\footnote{This is needed to ensure that the metric \eqref{eq:ansatz}
is $C^{\infty}$ at the center when expressed in Cartesian coordinates, but is also automatically implied by the field equations themselves.}.
Absence of an angle deficit at 
 $r=0$  would also require $ G=1$ there,
 but that condition can be imposed
 without loss of generality because the
 field equations are homogeneous in $G$,
 [i.e., as already mentioned, if $(F,G)$ is a solution, also 
  $(F,k G)$, with $k$ a constant rescaling factor,
  is a solution to the same theory].
 This ansatz implies, for $F$ and $\Gamma$, the functional form
\begin{subequations}\label{eq:UVSeries}
\begin{align}
&F\left(r\right) =  \sum_{n=0}^{\infty}  F_{2n+1} r^{2n+1}, \\
&\Gamma\left(r\right) =  \frac{1}{r} + \sum_{n=0}^{\infty}  \Gamma_{2n+1} r^{2n+1}.
\end{align}
\end{subequations}
Replacing this ansatz in the field equations \eqref{eq:GammaODEs} and \eqref{eq:GammaConstraint} we find that the coefficients \(F_{2n+1}\) and \(\Gamma_{2n+1}\) are all given in terms of two integration constants, $F_1$ and $\Gamma_1$. We have checked that the resulting perturbative solution leads to regular curvature invariants \(K_{ij}K^{ij}\), \(K\), \(R\) and \(R^{(3)}\) at the origin. 
Note that because we have fixed the center to be at $r=0$, in the numerical investigation
below we will not be allowed to use the shift symmetry [parameter $a$ in Eqs.~\eqref{eq:scaling}--\eqref{eq:scaling2}], whereas we will exploit the rescaling symmetry [parametrized by $b$ in Eqs.~\eqref{eq:scaling}--\eqref{eq:scaling2}].

The strategy is to use the perturbative solution \eqref{eq:UVSeries}, valid near the center, to provide initial data for $F$ and $\Gamma$ at some small radius $r\ll 1$, and integrate numerically outwards (once the integration constants $F_1,\, \Gamma_1$ have been chosen). Similarly, one can use the IR solution \eqref{eq:IR_series}
(fixing the integration constants $F_\infty,\, f_{(1,0)}$) to provide initial data at a finite radius  $r\gg 1$ for a numerical integration inwards. One then matches the two solutions smoothly at some fixed radius $r_m\sim 1$, where both solutions are regular, by imposing 
\begin{align}\label{eq:Jumps}
    \Delta F \rvert_{r_m} = \Delta F' \rvert_{r_m} = \Delta \Gamma \rvert_{ r_m} = \Delta \Gamma ' \rvert_{r_m} = 0,
\end{align}
where 
\begin{align}
    \left.\Delta X\right|_{r_m}=X_{\rm out}(r_m)-X_{\rm in}(r_m)
\end{align}
refers to the difference between the values of the function $X(r)$ when approaching the matching point \(r_m\) from the two directions. Equivalently, one can think of
this problem as that of finding the root(s) of the system  
\begin{align}\label{eq:RootFindingProblem}
\mathbf{F}\left(\mathbf{p}\right) = 0,    
\end{align}
where \(\mathbf{p}=\left(F_1, \Gamma_1, F_\infty, f_{(1,0)}\right)\) and the components of  \(\mathbf{F}\) are the jumps shown in \eqref{eq:Jumps}.

As mentioned earlier, the symmetry under rescaling of coordinates can still be used to eliminate one of the four integration constants [$F_\infty, f_{(1,0)}$ for the outer solution and $F_1,\Gamma_1$ for the inner one]. We choose for instance to set $F_1=1$. The system we have to solve is then overdetermined: we have four junction conditions \eqref{eq:Jumps} for three parameters. Thus, one does not expect existence of a regular solution on general grounds. 
To verify this, we consider three of the junction conditions given by Eq.~\eqref{eq:Jumps}, and we solve 
them (with a Newton-Raphson method)  in  our three variables $\Gamma_1$, $F_\infty$ and $f_{(1,0)}$. We then check whether the  fourth junction condition is satisfied (to within numerical errors) and it is not. 
We have checked that this result is stable against the choice of the initial guess of the Newton-Raphson algorithm.

Based on this overwhelming evidence, we can therefore conclude that there exist no regular solutions approaching at large radii the IR BHs that we identified previously, even if we allow for a conical defect at the center.
This result excludes option {\it (a)} 
outlined in the beginning of the section.

Let us now consider option {\it (b)}. In that case, the metric function $rG(r)$ must diverge at both $r=\pm \infty$, remaining finite and non-vanishing in between. This implies that
its logarithmic derivative $\Gamma(r)$ must change sign at finite $r = r_*$. However, this is impossible due to Eq.~\eqref{eq:GammaConstraint}. If $\Gamma(r_*) = 0$, the left-hand side of this equation becomes a sum of strictly negative terms, implying that also $F'(r_*)$ and $\Gamma'(r_*)$ must vanish. The latter means that $\Gamma(r)$ does not actually cross zero, and we arrive at a contradiction\footnote{One can be more careful and Taylor expand $\Gamma(r)$ in the vicinity of $r_*$ to see its behavior in more detail. One then obtains $\Gamma(r)\propto (r-r_*)^2$, which confirms that $\Gamma(r)$ does not change sign.}.

Option {\it (c)} still remains a logical possibility. 
We have not attempted to rule out robustly for generic values of $\lambda$, as we did with option {\it (a)}. However, given our experience in the structure of solutions to Eqs.~\eqref{eq:GammaODEs1}, \eqref{eq:GammaConstraint}, we believe it is unlikely. This is corroborated by the analysis in the limit $(\lambda-1)\ll 1$ presented in the next section.


\section{A perturbative expansion in $(\lambda - 1)$}\label{sec:LambdaExpansion}

Let us now give an additional analytic argument showing that no regular UV extension to our IR BH solutions exists. 
The interesting UV fixed point of the renormalization group flow \eqref{fixed_point} is close to $\lambda=1$ and, at least along some of the flow lines, $\lambda$ further approaches $1$ when the theory flows towards IR \cite{Barvinsky:2017kob}. This motivates to study the behavior of the solutions by performing a perturbative expansion in $\epsilon\equiv\lambda-1$.

According to standard techniques dealing with differential equations with small parameters in front of the highest derivatives, we introduce a rescaled coordinate \(\tilde r = r/\epsilon^{1/2}\). Recalling that $\sigma\approx (\lambda-1)/2$ in the desired limit, the asymptotics \eqref{eq:asymptoticsGamma} and \eqref{eq:IR_seriesGamma} for $\Gamma(r)$ suggest the following ansatz,
\begin{equation}
    \label{Gammatildeg}
\Gamma(\tilde r)=\frac{1+\epsilon\tilde g(\tilde r)}{\epsilon^{1/2}\, \tilde r},
\end{equation}
where the function $\tilde g$ is of order one and will interpolate between small and large $\tilde{r}$. We will see shortly that this ansatz provides the most general solution to the field equations \eqref{eq:GammaODEs1}, \eqref{eq:GammaConstraint} in the relevant limit $\epsilon\ll 1$. Notice that the divergence of $\Gamma(\tilde r)$ at $\tilde r=0$ excludes the wormhole {\it (b)} and throat {\it (c)} scenarios.

Substituting the ansatz into \(\mathcal{E}_1\) and \(\mathcal{E}_3\) and retaining only the leading terms in $\epsilon$, we obtain
\begin{subequations}\label{smalll}
\begin{align}
\label{smalll1}
& F''+\frac{ F'}{\tilde r}-\frac{F}{\tilde r^2}(1-\tilde r \tilde
g'-\tilde g)=0,\\
\label{smalll2}
&\frac{8\tilde g''}{\tilde r}-\frac{16\tilde g}{\tilde r^3}-2 FF'=0,
\end{align}
\end{subequations}
where now the derivatives are taken with respect to $\tilde r$. 
Note that, despite a lot of simplifications, this is still a system of two second-order differential equations, like the original system \eqref{eq:GammaODEs1}, \eqref{eq:GammaConstraint}. Its general solution contains four arbitrary integration constants, implying that we have not lost any solutions in making the ansatz \eqref{Gammatildeg}. 

Equation \eqref{smalll2} can be integrated once, yielding
\begin{equation}
    \label{tildeint}
8\bigg(\frac{\tilde g'}{\tilde r}+\frac{\tilde g}{\tilde r^2}\bigg)-F^2=A,
\end{equation}
where the integration constant $A$ must be fixed by suitable boundary conditions.
Since $\sigma\sim \epsilon/2$ as $\epsilon\to 0$, the asymptotics \eqref{eq:asymptotics}
correspond to the boundary conditions 
\begin{align}\label{smallbc}
    F\to F_\infty~,~~~~ \tilde
g\to 1~~~\text{at}~~ \tilde r\to +\infty,
\end{align}
from which one obtains $A=-F_\infty$.
We can use the scaling transformation [corresponding to the parameter $b$ in Eqs.~\eqref{eq:scalings}] to set $F_\infty=1$ and hence $A=-1$.
Moreover, from Eq.~\eqref{tildeint} we also obtain a sub-leading term in $F$,
\begin{align}\label{Fassmalll}
F(\tilde r)= 1+\frac{4}{\tilde r^2} + {\cal O}\left(\frac{1}{\tilde r^3}\right).
\end{align}

The combination of $\tilde g$ and its derivative in Eq.~\eqref{tildeint} is the same as in Eq.~\eqref{smalll1}. By combining the two equations one then obtains a closed second-order equation for $F(\tilde r)$
\begin{align}\label{smallF}
F''+\frac{F'}{\tilde r}+\bigg(\frac{F^2-1}{8}-\frac{1}{\tilde r^2}\bigg)F=0.
\end{align}
This is still a non-linear differential equation, which, to the best of our knowledge, cannot be solved analytically. Nevertheless, its numerical analysis is straightforward. Starting from large $\tilde r$ with the boundary conditions \eqref{Fassmalll} and integrating inwards, we find that $F$ diverges, producing a curvature singularity at the center. 

Alternatively, we can assume existence of a regular center. From the expansion  \eqref{eq:UVSeries} near $\tilde r=0$, it follows that the corresponding boundary conditions are
\begin{align}
    F\approx \tilde F_1 \tilde r,\quad \tilde g \approx \Gamma_1\tilde r^2\;,
\end{align}
where $\tilde F_1=\epsilon^{1/2} F_1$ and $\Gamma_1=-1/24$ is fixed from Eq.~\eqref{tildeint} 
by using the boundary condition at spatial infinity ($A=-1$). We have numerically integrated Eq.~\eqref{smallF} from $\tilde r=0$ with initial conditions $F(0)=0$, $F'(0)=\tilde F_1$ and scanned over different values of the single free parameter $\tilde F_1$. We have observed that the solution always oscillates at large $\tilde r$ around $1$ or $-1$ with a non-vanishing amplitude, and cannot be matched to the asymptotics \eqref{smallbc}. This once again rules out the possibility of a regular center inside the IR BH.

\section{Discussion} \label{sec:conclusions} 

Many puzzles of quantum gravity are related to BHs. To set up the stage for addressing these puzzles in a UV-complete theory, we looked for circularly symmetric stationary non-rotating vacuum solutions in $(2+1)$-dimensional Ho\v rava gravity. We found that in the presence of a positive cosmological constant the theory possesses, unlike $(2+1)$-dimensional GR, a family of solutions with two Killing horizons: the outer cosmological horizon, and the inner horizon that corresponds to a BH from the low-energy perspective. At large distances the solutions asymptotically approach de Sitter spacetime with a possible finite angle deficit. In the limit of vanishing cosmological constant the asymptotic spacetime is locally flat, but features a global growing (negative) deficit angle.

Motivated by the conjecture that the good quantum properties of Ho\v rava gravity may lead to resolution of BH singularities (see e.g. Ref.~\cite{Blas:2014aca,chojnacki2021finite}), we scrutinized the regularity of our BH solutions. We found that they are singular at the center, similar to BHs in GR, implying that no resolution of singularities occurs in the pure vacuum theory. Stated differently, we have ruled out the existence of regular classical solutions in pure $(2+1)$-dimensional Ho\v rava gravity (``gravitational solitons'') with BH-type Killing horizons.

Our results can have several interpretations. It can be that the BH solutions we found are merely physically irrelevant. A more interesting possibility is that they may form as the geometry
describing the exterior of collapsing matter configurations. In that case, the fate of the central singularity will depend on the dynamics of matter. For example, the latter can form a compact remnant inside the Killing horizon,  smoothing out the metric at the center. Alternatively, regular solutions may be dynamical (see e.g. Ref.~\cite{Izumi:2009ry, Mukohyama:2010xz}). For instance, matter can bounce back from the center, in which case the BH solution will correspond to transient configurations.\footnote{Such bounce is in principle classically allowed in Ho\v rava gravity, because the Killing horizon is not the true event horizon for high-energy modes that propagate with arbitrarily high velocities.}

In this context it is instructive to discuss the gravitational energy of the BH solutions. Recall first that we did not impose the global Hamiltonian constraint following from the variation of the action with respect to the lapse $N(t)$,
\[
\frac{\delta S}{\delta N}=0\quad\Rightarrow \quad
\int d^2x \sqrt{\g} \big(K_{ij}K^{ij}-\lambda K^2+\mu R^2+2\Lambda\big)=0.
\]
We find this constraint meaningless for spacetimes with non-compact spatial slices, like in our case: a positive energy in one region of space can be compensated by a negative contribution from another region infinitely far away. An alternative viewpoint is that we have studied the version of the theory where the lapse is set to $N=1$ from the start and there is no gauge freedom of time-reparametrization. Therefore, the theory possesses a well-defined notion of local and global energy, given by the Hamiltonian.  

Applying the Legendre transform to the Lagrangian \eqref{eq:action}, we find the Hamiltonian of pure Ho\v rava gravity,
\begin{align}\label{eq:hamiltonian}
    {\cal H}=&\frac{1}{\kappa}\int d^2 x \sqrt{\gamma}\  \left(K_{ij}K^{ij}-\lambda K^2 +\mu R^2 + 2 \Lambda \right) \nonumber\\
    & -\frac{2}{\kappa}\int d^2 x \sqrt{\gamma}\,
    \mathcal{P}_i N^i
    + \frac{1}{\kappa}\oint d\Sigma^i q_i   ,
\end{align}
where
\begin{align}
    q_i = 2 N_i \left( K^{ij} - \lambda K \gamma^{ij} \right),
\end{align}
and $d\Sigma^i$ denotes the line element vector on the boundary at spatial infinity. Notice that we do not include any York--Gibbons--Hawking term~\cite{Gibbons:1976ue, York:1972sj} neither in the action, nor in the Hamiltonian. This is justified, since the field equations are fourth-order in spatial derivatives of the metric $\gamma_{ij}$ and thus the variational principle requires fixing both $\delta\g_{ij}$ and its derivatives on the spatial boundary to zero. The variation of the action \eqref{eq:action} 
is then well-defined without any boundary term.
The Hamiltonian \eqref{eq:hamiltonian} does not include the contribution of matter, which, as we argued, must be considered in the full physical setup. However, we can use it to compute the  energy arising from the gravitational field outside matter configurations.

To simplify further discussion, let us set $\Lambda=0$. Then, for stationary solutions, like our BH metric, the gravitational energy can be cast into a boundary integral using the following relation,
\begin{multline}
    \sqrt{\gamma}\big(K_{ij}K^{ij}-\lambda K^2+\mu
    R^2\big)=\sqrt{\gamma}\,\gamma_{ij} {\cal G}^{ij}\\
    +(1-2\lambda)\partial_t(\sqrt\gamma\, K)
    +\sqrt{\gamma} \,\nabla_i l^i   ,
\end{multline}
where
\begin{align}
    l^i=(2\lambda-1)N^i K-2\mu\nabla^i R\; .
\end{align}
Thus, using Gauss law we can write
\begin{align} \label{eq:constraint_gauss}
   {\cal H}_{\rm tot}={\cal H}_{\rm out}
   +{\cal H}_{\rm center}+
   \mathcal{O}\left(\mathcal{P}_i, \mathcal{G}^{ij}\right),
\end{align}
where 
\be
{\cal H}_{\rm out}=\frac{1}{\kappa}\oint
d\Sigma^i (l_i + q_i )\;,
\ee
the term ${\cal H}_{\rm center}$ includes possible matter contribution in the central region, as well as the integral of $l_i$ over the line encompassing this region, and $\mathcal{O}\left(\mathcal{P}_i, \mathcal{G}^{ij}\right)$ denotes terms that vanish on shell (in vacuum and away from singularities). The long-distance contribution ${\cal H}_{\rm out}$ is evaluated using the asymptotics $F=F_{\infty}r^{-\sigma}$,
$G=\hat G_\infty r^{2\sigma}$ at $r\to\infty$ with the result
\begin{align}
{\cal H}_{\rm out}= \frac{2 \pi}{\kappa} F_\infty^2 \hat G_\infty\left(1+3 \sigma\right)\;.
\end{align}
We observe that this contribution is finite and positive. 

The finiteness of the BH gravitational energy is consistent with the proposal that this metric can form outside matter configurations during gravitational collapse. To investigate this possibility in more detail, one would need to follow the dynamics of time-dependent spherical collapse in this theory. Unlike in the infrared limit of Ho\v rava gravity, where gravitational collapse has been studied in several works~\cite{Garfinkle:2007bk,Bhattacharyya:2015uxt,Saravani:2013kva,Akhoury:2016mrc,Franchini:2021bpt}, numerical simulations  in the UV theory are complicated by the presence of higher (spatial) derivatives, which 
would require to carefully examine the character of the resulting (non-linear) system of partial differential equations, the well-posedness of the Cauchy problem, etc. Clearly, more work is  needed 
in this direction.

\section*{Acknowledgments}
This work is dedicated to the memory of Renaud Parentani, a wonderful person and a brilliant scientist whom we will all deeply miss. 
We thank S. Liberati and S. Solodukhin for insightful and illuminating discussions on Lorentz violating gravity and BH physics. Our work has been supported by the European Union's H2020 ERC Consolidator Grant ``GRavity from Astrophysical to Microscopic Scales'' grant agreement no. GRAMS-815673 (E. B., M. H-V. and G. L.) and the Russian
Foundation for Basic Research grant 20-02-00297 (S.S.).

\newpage


\appendix

\begin{widetext}
\section{\label{sec:ExplicitEOMs} Equations of motion}
In this Appendix, we provide explicit expressions for the field equations in terms
of the metric functions $F(r)$ and $G(r)$, using the same notation as in the main text. In more detail, the explicit expressions for Eqs.~\eqref{eq:eom} are 
\begin{align}
    E_1 & = (\lambda -1) \left(r^2 \left(G F' G'+G \left(G F''+F G''\right)-F \left(G'\right)^2\right)+G^2 r F'-F G^2\right)+F G
   r^2 G''+2 F G r G'   ,
\end{align}
\begin{align}
    E_2 =& (\lambda -1) \left[G r^3 \left(F \left(2 G \left(G F''+F G''\right)-F \left(G'\right)^2\right)+G^2 \left(F'\right)^2+4
   F G F' G'\right)+2 F G^2 r^2 \left(2 G F'+F G'\right)-F^2 G^3 r\right]\nonumber\\
   &+\mu  \left[4 r^3 \left(-4 \left(G'\right)^2 G''+4 G G^{(3)} G'+G \left(3 \left(G''\right)^2-2 G
   G^{(4)}\right)\right)-16 r^2 \left(-4 G G' G''+2 \left(G'\right)^3+G^2 G^{(3)}\right)\right.\nonumber\\
   &\left.+16 G r \left(2 G
   G''-\left(G'\right)^2\right)-32 G^2 G'\right]+2 G^3 r^3 \left(\left(F'\right)^2+F
   F''\right)
   -2 G^3 r^3\Lambda   ,
\end{align} 
while the constraint \eqref{eq:Constraint} is
\begin{align}
   E_3 =&  (\lambda -1) \left[-2 F G^2 r^2 \left(G F'+F G'\right)-G r^3 \left(G F'+F G'\right)^2-F^2 G^3 r\right]\nonumber\\
   &+\mu  \left[r^3 \left(-8 \left(G'\right)^2 G''+8 G G^{(3)} G'-4 G \left(G''\right)^2\right)+8
   r^2 \left(-G G' G''-2 \left(G'\right)^3+G^2 G^{(3)}\right)+16 G r \left(G G''-3 \left(G'\right)^2\right)\right.\nonumber\\
   &\left.-16 G^2
   G'\right]-2 F G^3 r^2
   F'-2 F G^2 r^3 F' G'   
  +2 G^3 r^3\Lambda  .
\end{align}
%
\end{widetext}

\section{\label{sec:WKB} Linearized analysis  at large radius}

To study the asymptotics of the general solution of the system \(\left(\mathcal{E}_1,
\mathcal{E}_3\right) \) at large $r$, we substitute the expansion \eqref{asexpand}
and linearize, assuming the functions $f,g$ and their first derivatives
are small ($f,g\ll 1$ and $r f',r g'\ll 1$). We do not need to make any assumptions
about the second derivatives of $f$ and $g$. Proceeding in this way, we obtain the system of linear equations
\begin{subequations}\label{linas}
\begin{align}
\label{linas1}
    (\lambda-1)f'' &+\frac{(\lambda-1)}{r} f'+
    \frac{\lambda(1+2\sigma)}{r}g'
    +\\
    &+\frac{(1+2\sigma)^2-(\lambda-1)\sigma(\sigma+1)}{r^2}g=0,
    \nonumber\\
    8(1+2\sigma)^2g''&+\frac{16\sigma(1+2\sigma)^2}{r}g'\nonumber\\
    &-2(\lambda+\sigma+\lambda\sigma)
    F_\infty^2 r^{1-2\sigma}f'\nonumber\\
    &-(\lambda-1)(1+4\sigma+3\sigma^2)
    F_\infty^2 r^{-2\sigma} g \nonumber\\  
    &=\frac{16\sigma(1+2\sigma)^2(2+\sigma)}{r^2}.
\label{linas2}
\end{align}
\end{subequations}
Note that in deriving Eq.~\eqref{linas} we have assumed that $\sigma<1$ and neglected terms of order
$O(r^{-2}g)$ in Eq.~\eqref{linas2}, which are small compared to the terms we have kept. 

We are now interested in the solutions of this system at large $r\gg 1$.
A particular solution is provided by 
\begin{align}
    f&=f_{\infty} r^{-2(1-\sigma)}, &
    g&=g_{\infty} r^{-2(1-\sigma)},
\end{align}
where the coefficients $f_{\infty}$, $g_{\infty}$ are determined from the
linear algebraic equations,
\begin{align}
    &4(\lambda-1)(1-\sigma)^2 f_\infty+\nonumber\\
    &+(1-2\lambda+5\sigma-3\lambda\sigma
    +5\sigma^2+3\lambda\sigma^2)g_\infty=0,\\
    &4(\lambda+\sigma+\lambda\sigma)(1-\sigma)f_\infty+\nonumber\\
    &-(\lambda-1)(1+4\sigma+3\sigma^2)g_\infty=\frac{16\sigma(1+2\sigma)^2(2+\sigma)}{F_\infty^2}.
\end{align}
In particular, at $\lambda-1\ll 1$ we have $f_{\infty}=4(\lambda-1)/F_\infty^2$,
$g_\infty=0$. 

It is straightforward to see that two solutions of the homogeneous
system at large $r$ have the form
\begin{align}
    f &=\frac{\sigma}{r}, & g&=\frac{1}{r} , &\text{ and } \quad
    f &=\text{const}, & g&=0 .
\end{align}
As discussed in the main text, they correspond to the shift and
rescaling of the non-linear solution.

The two other solutions cannot in general be found
analytically. Still, they can be derived in the limit $(\lambda-1)\ll
1$. To simplify the subsequent analysis, we set $F_\infty=1$, as can always be achieved by the symmetry transformation~\eqref{eq:scalings}. 

We observe that the highest derivative term 
in Eq.~\eqref{linas1} is multiplied by a small quantity, which suggests
using the Wentzel--Kramers--Brillouin (WKB) expansion \cite{Bender:1977}. The derivatives of a function should be
treated as enhanced by a factor $1/\sqrt{\lambda-1}$ compared to the
function itself. We will need both the leading and the subleading
terms in the expansion in powers of $\sqrt{\lambda-1}$. With this in
mind and recalling that $\sigma\approx  (\lambda-1)/2$, we simplify Eqs.~\eqref{linas}, keeping only the relevant terms:
\begin{subequations}\label{linasWKB}
\begin{align}
\label{linasWKB1}
&(\lambda-1)f''+\frac{\lambda-1}{r}f'
+\frac{g'}{r}+\frac{g}{r^2}=0 , \\
 \label{linasWKB2}
 &8g''-2 r ^{1-2\sigma}f'=0
\end{align}
\end{subequations}
The form of the equations suggests the following ansatz,
\begin{subequations}\label{WKBfg}
\begin{align}
    &f=(f_0+\sqrt{\lambda-1}f_1+\ldots)\exp\bigg(\frac{iQ}{\sqrt{\lambda-1}}\bigg),\\
    &g=\sqrt{\lambda-1}\,
    (g_0+\sqrt{\lambda-1}g_1+\ldots)\exp\bigg(\frac{iQ}{\sqrt{\lambda-1}}\bigg),
\end{align}
\end{subequations}
We first consider the leading order, which corresponds to terms $O(1)$
and $O(1/\sqrt{\lambda-1})$
in Eqs.~\eqref{linasWKB1} and \eqref{linasWKB2}, respectively. At this
order, we obtain
\begin{subequations}
\begin{align}
&-(Q')^2f_0+\frac{i Q'}{r}g_0=0,\\
&-2ir^{1-2\sigma}Q'f_0-8(Q')^2g_0=0.
\end{align}
\end{subequations}
As $f_0$, $g_0$ are non-vanishing by assumption, the system must be
degenerate, giving the condition
\begin{equation}
\label{QWKB}
    (Q')^2=\frac{r^{-2\sigma}}{4}.
\end{equation}
Choosing the positive root, we find $Q= r^{1-2\sigma}/2$ and
\begin{equation}
    \label{WKBg0f0}
g_0=-\frac{i}{2}r^{1-\sigma}f_0\,.
\end{equation}
The fact that $Q$ is real implies that the solution is quickly
oscillating. 

To find the behavior of the amplitude, we need to go to
the next WKB order. This corresponds to terms $O(\sqrt{\lambda-1})$ in
Eq.~\eqref{linasWKB1} and $O(1)$ in Eq.~\eqref{linasWKB2}. Using that
$Q''=O(\lambda-1)$, we obtain
\begin{subequations}
\begin{align}
    &2iQ'f_0'+\frac{iQ'}{r}f_0
    +\frac{g_0'}{r}
+\frac{g_0}{r^2}
-(Q')^2f_1+\frac{iQ'}{r}g_1=0,\\
&16iQ'g_0'-2r^{1-2\sigma}f_0'
-2iQ'r^{1-2\sigma}f_1-8(Q')^2g_1=0
\end{align}
\end{subequations}
Next, we multiply the second equation by $iQ' r^{-1+2\sigma}/2$ and
add it to the first one. This eliminates the functions $f_1$ and $g_1$,
so that we are left with an equation containing only $f_0$ and
$g_0$. Using further the relation \eqref{WKBg0f0}, we obtain a
differential equation for $f_0$,
\begin{equation}
    f_0'+\frac{(1-\sigma)f_0}{2r}=0.
\end{equation}
We then obtain that \(f_0\propto\ r^{-(1-\sigma)/2}\) and $g_0\propto  r^{(1-\sigma)/2}$, where the latter is a growing function of the radial coordinate. 

The above analysis shows that for $(\lambda-1)\ll 1$ two linearly independent solutions of the system \eqref{linas} oscillate with a growing amplitude. By solving the system numerically we have found that this qualitative behavior persists at finite $(\lambda-1)$ as long as $\lambda\lesssim 5/2$. 

Let us make the following comment. At first sight, it may be surprising to find oscillatory asymptotics in stationary perturbations of a time-independent background. Normally, one would expect such perturbations to obey an elliptic equation, which should lead to solutions that exponentially grow or decay at large $r$. The fact that the perturbations are instead oscillating in $r$ seems to suggest that the spatial part of the eigenmode equation in the BH background has turned hyperbolic, and one may worry if this leads to a rapid gradient instability when the time evolution is included. In more detail, the WKB result \eqref{QWKB} could suggest that the dispersion relation for the short-wavelength modes at small $(\lambda-1)$ has changed from \eqref{dispflat} to
\begin{equation}
\label{wrongdisp}
\omega^2=4\mu(\l-1)\, k^4- F^2(r)\, k^2\;,   
\end{equation}
where we have used that $r^{-\sigma}=F(r)$. This would imply an instability on time scale
$\tau_{\rm inst}\sim F/\sqrt{\mu(\lambda-1)}$, which would be catastrophic.

Fortunately, this is not the case. The reason is that our BH background is stationary, rather than static, i.e. it has non-vanishing shift vector. Therefore, the time derivative operator in any field equations gets modified by an admixture of a term with spatial derivatives (cf. Eq.~\eqref{Lieder}), $\d_t\mapsto \d_t-N^i\d_i+\ldots$, where dots stand for term without derivatives acting on the field. As a consequence, the dispersion relation for short-wavelength modes takes the form
\be
\big(\omega-F(r)\,k\big)^2=4\mu (\lambda-1) k^4\;,
\ee
which for $\omega=0$ is the same as Eq.~\eqref{wrongdisp}. But now $\omega$ never becomes imaginary, and no catastrophic instabilities develop. Notice that this does not prove the absence of long-wavelength instabilities that are, anyway, less harmful.

\bibliography{bh_HG}{}

\end{document}